\documentclass[journal,twoside,web]{ieeecolor}
\usepackage{generic}
\usepackage{cite}
\usepackage{amsmath,amssymb,amsfonts}
\usepackage{algorithmic}
\usepackage{graphicx}
\usepackage{textcomp}
\usepackage{bm}
\usepackage{multirow}
\usepackage{booktabs}
\usepackage{hyperref}

\def\BibTeX{{\rm B\kern-.05em{\sc i\kern-.025em b}\kern-.08em
    T\kern-.1667em\lower.7ex\hbox{E}\kern-.125emX}}
\begin{document}
\title{Spatio-Temporal Classification of Lung Ventilation Patterns using 3D EIT Images: A General Approach for Individualized Lung Function Evaluation}

\author{Shuzhe Chen, Li Li, Zhichao Lin, Ke Zhang, Ying Gong, Lu Wang, Xu Wu, Maokun Li, \IEEEmembership{Senior Member, IEEE}, Yuanlin Song, Fan Yang, \IEEEmembership{Fellow, IEEE}, and Shenheng Xu, \IEEEmembership{Member, IEEE},
\thanks{This paragraph of the first footnote will contain the date on 
which you submitted your paper for review. This work was supported in part by the Institute for Precision Medicine,  
Tsinghua University, National Natural Science Foundation of China (61971263 and 62171259), 
Biren Technology, and BGP Inc. Shuzhe Chen and Li Li contributed equally to this work. \emph{(Corresponding author: Maokun Li.)}}
\thanks{Shuzhe Chen, Zhichao Lin, Ke Zhang, Maokun Li, Fan Yang and Shenheng Xu are with Department of Electronic Engineering, Institute of Precision Medicine, Tsinghua University, Beijing 100084, China, Beijing National Research Center for Information Science and Technology (BNRist),  (e-mails: csz21@mails.tsinghua.edu.cn, lzc19@mails.tsinghua.edu.cn, kzhang320@mail.tsinghua.edu.cn, maokunli@tsinghua.edu.cn, fanyang@tsinghua.edu.cn, shxu@tsinghua.edu.cn).}
\thanks{Li Li, Ying Gong, Lu Wang, Xu Wu, and Yuanlin Song are with the Department of Pulmonary and Critical Care Medicine, Zhongshan Hospital, Fudan University, 180 Fenglin Rd, Shanghai 200032, China (e-mails:li.li@zs-hospital.sh.cn, gong.ying@zs-hospital.sh.cn, wu.xu@zs-hospital.sh.cn, bluewang723@163.com, song.yuanlin@zs-hospital.sh.cn).}
}
\maketitle

\begin{abstract}
The Pulmonary Function Test (PFT) is an  widely utilized and rigorous classification test for lung function evaluation, serving as a comprehensive tool for lung diagnosis. Meanwhile, Electrical Impedance Tomography (EIT) is a rapidly advancing clinical technique that visualizes conductivity distribution induced by ventilation. EIT provides additional spatial and temporal information on lung ventilation beyond traditional PFT. However, relying solely on conventional isolated interpretations of PFT results and EIT images overlooks the continuous dynamic aspects of lung ventilation.
This study aims to classify lung ventilation patterns by extracting spatial and temporal features from the 3D EIT image series.
The study uses a Variational Autoencoder network with a MultiRes block to compress the spatial distribution in a 3D image into a one-dimensional vector. These vectors are then concatenated to create a feature map for the exhibition of temporal features. A simple convolutional neural network is used for classification.
Data collected from 137 subjects were finally used for training. The model is validated by ten-fold and leave-one-out cross-validation first. The accuracy and sensitivity of normal ventilation mode are 0.95 and 1.00, and the f1-score is 0.94. Furthermore, we check the reliability and feasibility of the proposed pipeline by testing it on newly recruited nine subjects. Our results show that the pipeline correctly predicts the ventilation mode of 8 out of 9 subjects.
The study demonstrates the potential of using image series for lung ventilation mode classification, providing a feasible method for patient prescreening and presenting an alternative form of PFT.
\end{abstract}


\section{Introduction}
\label{sec:introduction}
\IEEEPARstart{R}{espiration} diseases are the third leading cause of death worldwide and severely impact people's quality of life\cite{fei2023symptom}. The prevalence of chronic respiratory diseases (CRDs) has increased by about 40\% in the past thirty years\cite{lorenzo2020prevalence}. However, the diagnosis rate of CRDs is far lower than the prevalence rate, and the treatment rate is even lower than the diagnosis rate\cite{stolz2022towards, jin2022screening}. Early screening and diagnosis of ventilation diseases are of critical significance, yet it has not received enough attention.

The lung ventilation function is evaluated using the Pulmonary Function Test (PFT), in which the airflow inhaled and exhaled by the lungs is recorded and measured by a flow meter. A well-trained physician instructs the subjects to alternate between tidal breath and forced expiration following the regulations of the American Thoracic Society (ATS)/European Respiratory Society (ERS)\cite{stanojevic2022ers}.
Like other clinical tests, such as blood tests, PFT requires establishing normal values for accurate diagnosis. However, the challenge with PFT is that normal values can vary significantly from person to person\cite{choi2005normal} compared to other tests. Ventilation performance depends on numerous factors, including age, gender, body mass index, and even geography\cite{hegewald2021impact, brandli1996lung}.

Multivariate regression based on PFT results of many normal subjects is used to establish predicted values for a specific person. However, recruiting and measuring enough normal people in a specific area is challenging and demanding. Moreover, PFT results can only evaluate lung function without providing spatial information.
The large-scale implementation of PFT is challenging for several reasons. First, it requires a significant workforce and material resources, making it difficult to carry out in underdeveloped areas. Second, the testing cycle is long, and data statistics and analysis are highly demanding. Additionally, the lung function of the population changes objectively and dynamically over time \cite{talman2021pulmonary}, making it difficult to carry out periodic repetitions. Nevertheless, industrialization has contributed to an escalation in environmental pollution, decreasing the number of available healthy individuals. This decline poses a challenge when attempting to recruit sufficient healthy subjects for research or studies \cite{zhu2012}.

Electrical Impedance Tomography (EIT) is an emerging medical imaging modality that can detect conductivity changes in the measured area. The air content in the lungs varies during ventilation, and the resulting changes, especially the spatial distribution of electrical conductivity, can be captured by electrodes placed around the chest\cite{brown2003electrical}. EIT images are of significant clinical value\cite{ sang2020narrative, frerichs2000electrical}, including Positive End-expiratory Pressure (PEEP) titration guidance\cite{franchineau2017bedside}, regional distribution of ventilation\cite{frerichs2016regional}, and ventilation and perfusion matching\cite{spinelli2021unmatched,liu2022prone}.

Compared to PFT, which provides an overall evaluation of lung function, EIT can discriminate whether changes in global lung function stem from alterations in ventilation distribution or variations in ventilation magnitude. This distinction enables a more precise assessment of regional lung function and facilitates targeted prescreening effort\cite{frerichs2017chest}. Typically, lung functions are assessed both before and after medical procedures. In this process, 2D images taken at specific time points are often compared and analyzed based on manually crafted features relying on prior physiological knowledge. More attention should be paid to considering the image series and drawing broad conclusions regarding lung function.

In this study, we use 3D EIT image series to classify lung ventilation function modes in a general view. We extract spatial-temporal information using a Variational Autoencoder (VAE). Our proposed method achieved the highest accuracy and AUC of 95.6\% and 0.96, respectively, in binary classification based on in-vivo measurements. Furthermore, we also applied this method to four-category problems (normal, restrict, obstruct, and mix), and the maximum accuracy and f1-score were 86.3\% and 0.90, respectively.
Our contributions are as follows:

\begin{itemize}	
	\item Three-dimensional lung ventilation image series are reconstructed in a low-cost, radiation-free, and non-invasive manner by EIT.
	\item Spatial and temporal information in 3D image sequences during forced exhalation were considered simultaneously, representing an improvement over previous analyses of isolated 2D images.
	\item A concise and practical VAE network has been proposed for dimensional image reduction.
	\item A new pipeline has been developed to classify lung ventilation patterns, aiming to facilitate lung function diagnosis without dependence on expensive predicted values.
	\item Classification of ventilation patterns is focused rather than changes in lung ventilation in a specific situation.
\end{itemize}

The remainder of the article is organized as follows. Section \ref{secA} examines related works on lung ventilation pattern classification and the clinical application of EIT. Section \ref{pre} provides a detailed introduction to the preliminaries of EIT and VAE. In Section \ref{method}, we describe the proposed method. The EIT measurement system and study protocol are presented in Section \ref{exp}. In Section \ref{dis}, we test and optimize the network's performance. Finally, the work is summarized in Section \ref{conclu}.

\section{Related Work}
\label{secA}
\subsection{Automated PFT Assisted by Machine Learning }
PFT is an established and effective diagnostic tool for assessing lung function. During the test, subjects are instructed to perform tidal breath and forced expiration, allowing for measurement of volume and speed. The results are typically interpreted by physicians using predefined cutoffs in accordance with published guidelines\cite{pellegrino2005interpretative} to identify a typical pattern. However, this process heavily relies on the doctor's experience and subjective judgment\cite{topalovic2019artificial}. Additionally, many people with chronic obstructive pulmonary disease (COPD) symptoms do not meet the diagnostic criteria\cite{keener2020redefining,miller2011interpreting}. PFT results are usually interpreted by clinicians using discrete numeric data according to published guidelines. However, inter-rater variability among clinicians is known to occur, and inaccuracies in interpretation can impact patient care. As a result, many studies have focused on developing automated interpretation systems based on PFT values to reduce misdiagnosis rates and alleviate the burden on doctors.

Automated interpretation of PFT values has been proposed as the first stage in modeling the decision-making process of physicians\cite{topalovic2017automated, delclaux2019no, topalovic2019artificial,5478898}. Moreover, more advanced classification methods have been developed. In \cite{bhattacharjee2022classification}, a multi-layer perceptron was proposed to classify obstructive and non-obstructive patients, achieving an accuracy of 83.7\% with the spirometry data of Forced Vital Capacity (FVC), Forced Expiratory Volume in one second (FEV$_{1}$), and Forced Expiratory Flow (FEF$_{25-75}$). Disease-specific prediction of COPD and DPLD, as obstructive and non-obstructive, respectively, achieved approximately 90\% accuracy in the training dataset.

In addition, researchers in \cite{ioachimescu2020alternative} have considered the area under the expiratory flow-volume curve as a new indicator, improving the diagnostic classification rates. Furthermore, PFT values are believed to contain adequate information currently neglected by the diagnostic workflow. A fully convolutional network was applied to extract this latent information from the sequence of Flow-Volume loop data in PFT \cite{bodduluri2020deep}, enabling discrimination of the structural phenotype of chronic obstructive pulmonary disease that traditional PFT interpretation cannot accomplish using discrete single values.

\subsection{Structure-based prediction of lung function}
The static structure of the lungs determines their dynamic function. Modalities such as CT, MRI, and X-ray have been used to assess lung volume, parenchymal change, airway structure, air-trapping, and other structural features. These features are believed to be able to predict the functional parameters of the lungs.

In \cite{park2023deep}, an end-to-end scheme was used to predict PFT values, including FVC and $\text{FEV}_{1}$, using low-dose chest CT images, achieving an accuracy of 89.6\% and 85.9\%. In \cite{westcott2019chronic}, lung ventilation heterogeneity in COPD patients was predicted using support vector machines (SVM) based on CT texture analysis. The PFT results and $^{3}$He-MRI were considered ground truth, and the predicted ventilation maps had an accuracy of 88\% and an AUC of 0.82.

An integrated 3D-CNN and parametric-response mapping model\cite{ho20213d} is proposed to classify COPD subjects using CT-based variables, achieving an accuracy of 89.3\% and a sensitivity of 88.3\% in five-fold cross-validation. 
Deep learning has also been applied to discover subvisual abnormalities in CT scans related to COVID-19 in an interpretable manner \cite{zhou2022interpretable}.

X-ray images were also used to get an early assessment of the lung function of coronavirus patients with the help of invariant markers\cite{elsharkawy2021early}. MRI-derived regional flow-volume loops were also applied to detect chronic lung allograft dysfunction in early-stage\cite{moher2019mri}. Meanwhile, dynamic and functional imaging develops fast among traditional modalities, 4D-CT \cite{yamamoto2014pulmonary,yamamoto2023four} and hyperpolarized MRI \cite{doganay2019time,voskrebenzev2018feasibility} are the typical representation.

\subsection{Clinical Application of EIT}

EIT is a non-invasive imaging technique that provides real-time information on the distribution of electrical conductivity changes in the lung tissue, which is directly related to the respiration phase. Unlike other imaging modalities, EIT does not involve ionizing radiation and is relatively inexpensive, portable, and can be used at the bedside. Therefore, it has great potential for monitoring and evaluating lung function in various clinical scenarios\cite{adler2012whither}. 

2D-EIT was studied on 14 healthy individuals \cite{9871104},which showed an accuracy of 98\% in predicting PFT values using EIT values alone. The device used in the study was portable, which may have the potential for home lung monitoring. Similarly, other studies have been conducted on 35 children with cystic fibrosis\cite{muller2018evaluation} and seven healthy individuals conducting forced expiration maneuvers\cite{ngo2016linearity}.

Regional lung function evaluation is crucial in clinical settings \cite{dubsky2015imaging}, but it is challenging to assess without EIT. Medical hypotheses that are based on logical deduction can be validated by EIT through the distribution of ventilation. Studies have reported the observation of spatial and temporal inhomogeneous ventilation distributions in patients with chronic obstructive pulmonary disease (COPD) \cite{lasarow2021regional,frerichs2021spatial,vogt2016regional,vogt2012spatial}, cystic fibrosis \cite{lehmann2016global}, idiopathic pulmonary fibrosis\cite{krauss2021evaluation}, and smokers  \cite{vogt2019regional}. These findings demonstrate the potential of EIT to provide valuable information on regional lung function in various respiratory conditions.

The wide acceptance of these physiological findings within the medical community confirms EIT's reliability. Consequently, once the hypothesis is validated, EIT has proven to be a valuable tool for assessing the effectiveness of different treatments. For example, studies have shown that EIT can be used to evaluate the effectiveness of pulmonary rehabilitation\cite{ma2022pulmonary, eimer2021effect}, bronchodilators\cite{frerichs2016regional}, position changes from bed to a wheelchair\cite{yuan2021effect}, and even cardiac surgery\cite{krause2014monitoring}. EIT can also guide tracheal tube placement\cite{lumb2018observational,lumb2020effects} and aid in PEEP titration\cite{ren2022comparison}.

Although EIT has shown promising results in correlating with PFT results and evaluating lung function in specific disease scenarios, there are still limitations in the current studies:
\begin{itemize}
	\item The sample size in most studies is relatively small, limiting the generalizability of the findings.
	\item There is still a lack of standardization in the EIT data acquisition and analysis procedures.
	\item EIT measurements are sensitive to various factors, such as electrode positioning, patient movement, and chest wall abnormalities, which may affect the accuracy and reproducibility of the results.
	\item EIT is still considered an emerging technology, and there is a need for further research to establish its clinical utility and cost-effectiveness in routine clinical practice.
\end{itemize}
Despite these limitations, EIT holds good promise in pulmonary medicine as a non-invasive, radiation-free, and portable lung function assessment and monitoring tool. Further studies with larger sample sizes and standardization of EIT data acquisition and analysis procedures are needed to establish its clinical usefulness.

\section{PRELIMINARIES}
\label{pre}
\subsection{EIT Formulation}
\label{EIT}
\begin{figure}[!h]
	\centering
	\includegraphics[width=90mm]{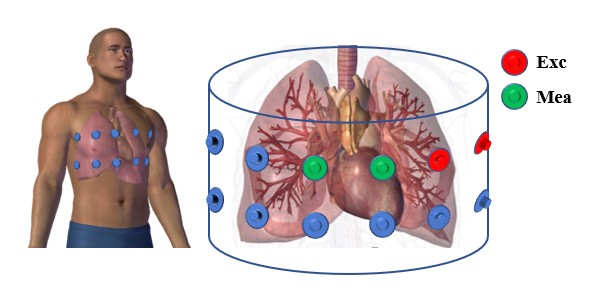}
	\caption{EIT imaging principle: Excitation electrodes (red) and measurement electrodes (green) is placed on the surface of the domain of interest (DOI) to apply small electrical currents and measure the resulting voltage distribution for image reconstruction.(The three-dimensional lung image is sourced from Zygote Media Group.)}
	\label{mea}
\end{figure}

EIT is a non-invasive medical imaging technique that utilizes small electrical currents to create images of the internal conductivity distribution within an object. In the context of lung imaging, EIT involves the placement of multiple electrodes on the surface of the chest. These electrodes serve as excitation points for applying small alternating currents. By measuring the resulting voltage distribution at each electrode, EIT can generate images depicting the lungs' internal conductivity distribution (see Fig.\ref{mea}). The conductivity distribution within the lungs changes as air is inhaled and exhaled during breathing. This dynamic feature of EIT makes it a valuable tool for monitoring lung function and detecting abnormalities in real-time. Moreover, EIT is radiation-free and non-invasive, making it a safe imaging technique for repeated and continuous monitoring of lung function.

Reconstructing the conductivity distribution from the boundary voltage data using EIT is a highly ill-posed problem \cite{scherzer2010handbook}, particularly in the case of 3D EIT. The algorithm used in this work is based on 3D time-difference image reconstruction, as described in previous studies\cite{zhang2019three,zhang2022deep}. The basic principle of this algorithm is briefly introduced below. The chest's conductivity distribution, denoted by $\bm \sigma$, is defined on a tetrahedral mesh. Let $\bm d$ and $\bm d_* $ denote the simulated and measured boundary voltage, respectively, given the injection-measurement pattern. The function $\bm S(*)$, also known as the forward model, maps the conductivity distribution to the boundary voltage. At an initialized point $\bm \sigma_{0}$, a linear approximation is needed:
\begin{equation}\label{eq1}
	\bm S({\bm \sigma}) \approx \bm S({\bm \sigma_{0}}) + \bm J{\bm \sigma_{0}}^T \cdot \Delta \bm \sigma
\end{equation}
where $\Delta \bm \sigma$ is a small conductivity change around $\bm \sigma_{0}$, and $\bm J_{\bm \sigma_{0}}$ is the Jacobian matrix of $\bm S({\bm \sigma})$ evaluated at $\bm \sigma_{0}$:
\begin{equation}\label{eq2}
	\bm J_{\bm \sigma_{0}} = \frac{\partial \bm S({\bm \sigma _0})}{\partial \bm \sigma _0}.
\end{equation}

Here, we focus on the difference in the conductivity distribution. Let the measured data and conductivity distribution at $t_{1}$ and $t_{2}$ be denoted by $\bm d_{1*}$, $\bm d_{2*}$ and $\bm \sigma_{1}$, $\bm \sigma_{2}$, respectively. The time-difference EIT can be approximated as follows:

\begin{equation}\label{eq3}
	\bm d_{2*}-\bm d_{1*} \approx \bm J_{\bm \sigma _0} \cdot(\bm \sigma_{2}-\bm \sigma_{1})
\end{equation}

\subsection{the Proposed VAE}
Variational autoencoder (VAE) is a deep generative network \cite{kingma2013auto} that can encode high-dimensional data into a lower-dimensional latent space representation. Herein, VAE is employed to acquire a compressed representation of 3D EIT images. The 3D EIT images are fed into the encoder network, which maps the data to a lower-dimensional latent code $\bm z$. 
\begin{figure}[!h]
	\centering
	\includegraphics[width=90mm]{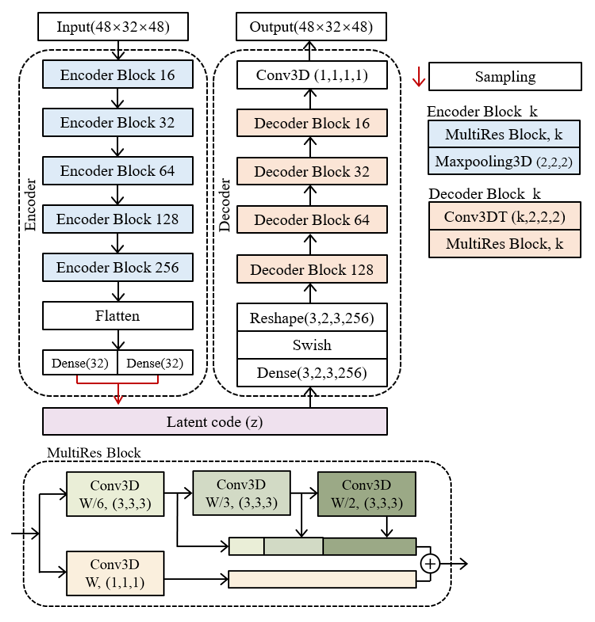}
	\caption{The VAE Workflow and Structure. The encoder network maps the input images to the latent variable distribution, which is then sampled to produce a latent variable. The decoder network takes the latent variable as input and reconstructs the original data.}
	\label{flow}
\end{figure}
Then a decoder would reconstruct the original data from $\bm z$ as shown in Fig.\ref{flow}. 

A three-dimensional convolutional neural network (CNN) is applied for both the encoder and decoder blocks. The encoder is stacked with five encoding blocks with different numbers of channels, followed by a flatten layer and a dense layer. The flattened output is then passed through two dense layers to obtain the mean and standard deviation of the latent space distribution. The mean and standard deviation are used to sample from the distribution and obtain the latent representation of the input image. The decoder consists of several layers of 3D transposed convolutional blocks, which perform the opposite operation of the encoder. The transposed convolutional blocks gradually increase the dimensions of the input until the output matches the original input dimensions.

A MultiRes block\cite{ibtehaz2020multiresunet} is the common part of the encoder and decoder and the key of the proposed structure. Three convolutional blocks are connected sequentially to capture spatial features at multiple resolutions. Moreover, a residual connection is introduced by a $1 \times 1$ convolutional layer to comprehend some additional spatial information. 

The purpose of VAE is to model the real probability distribution of the training data $p_{r}(\bm x)$ with a latent distribution $p(\bm z)$, which is commonly set as standard distribution $\mathcal{N}(0, \bm I)$. The distribution of the generated samples $p(\bm x)$ can be written as: 
\begin{equation}
	p(\bm x) = \int p(\bm x|\bm z) p(\bm z) \mathrm{d}\bm z
\end{equation}

Given the Kullback–Leibler (KL) divergence, which is a measure of the difference between the distributions. The optimization principle of VAE can be written as follows:
\begin{equation}
	\mathop{\arg\min}\limits_{{p(\bm x|\bm z)} }  D_{KL}(\left(p_{r}(\bm x)\|p(\bm x)\right))
\end{equation}
The corresponding objective function is derived as:
\begin{equation}
	\mathcal{L} = \mathbb{E}_{x\sim\tilde{p}(\bm x)}[D_{KL}(\left(q(\bm z|\bm x)\|p(\bm z)\right)) - \int q(\bm z|\bm x) \ln p(\bm x|\bm z) \mathrm{d}\bm z] 
	\label{KL}
\end{equation}

Assume that $q(\bm z|\bm x)\sim \mathcal{N}(\bm \mu, \bm \sigma)$, where $\bm \mu$ and $\bm \sigma = \rm diag \{ \bm \sigma_{v}^2 \} $
are the mean vector and variance vector of the VAE encoder. Then the first KL term in Eq.\ref{KL} can be written as:

\begin{equation}
	\mathcal{L}_{KL} = \frac{1}{2}\sum_{i=1}^{n}(\mu_{i}^{2}(x)+\sigma_{i}^{2}(x)-\ln\sigma_{i}^{2}(x)-1)
\end{equation}
where $n$ is the length of the latent code $z$. The KL loss encourages the distribution in the latent space to be close to a standard normal distribution, which leads to a smoother and more continuous latent space structure. The second term is approximated \cite{zhang2022deep, lin2022feature} as:

\begin{equation}
	\mathcal{L}_{MSE} = \frac{1}{N_{x}}\|\bm x - \hat{\bm x}\|_2^{2}
\end{equation}
where $\bm x$ and $\hat{\bm x}$ represent the raw and reconstructed images respectively, and $N_{x}$ denotes the total number of pixels in a single 3D image. This term ensures that the reconstructed images are faithful to the input images.

In the total loss, the $\mathcal{L}_{KL}$ is weighted by  $\lambda = 10^{-3}$ to prevent it from dominating the reconstruction loss during training. 

\begin{equation}
	Loss = 	\mathcal{L}_{MSE}+\lambda \mathcal{L}_{KL}
\end{equation}

The VAE is designed to learn unsupervised representations by extracting latent features with the encoder. These learned features can then be used as inputs to classification tasks, achieving improved performance.

\section{THE PROPOSED APPROACH}
\label{method}

\subsection{Voltage data preprocessing}
\begin{figure}[!h] 
	\centering
	\includegraphics[width=88mm]{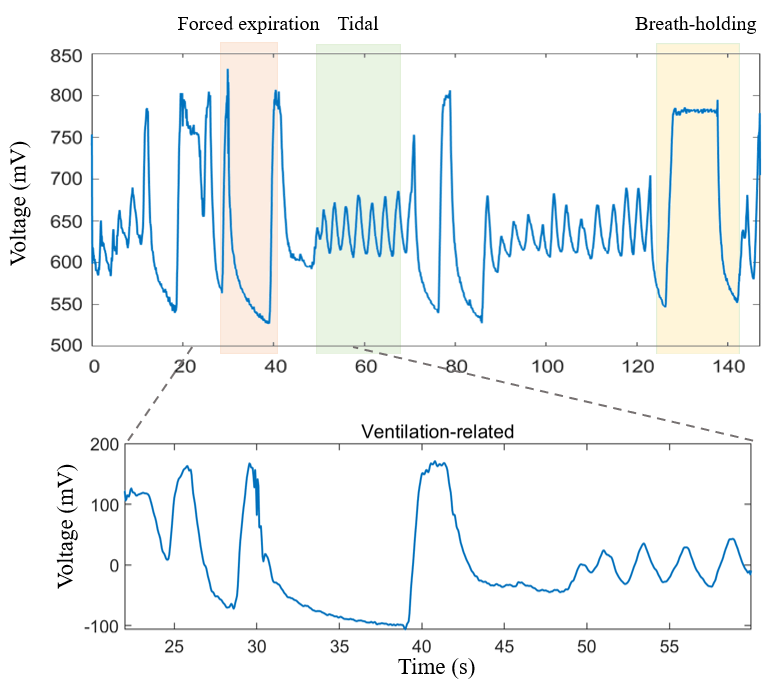}
	\caption{The raw EIT voltage signal throughout a whole PFT, and the separated ventilation-related signals around the forced expiration.}
	\label{raw}
\end{figure}

The total voltage signal obtained from EIT during a PFT is presented in Figure \ref{raw}. The signal captures the dynamics of ventilation, including forced expiration (depicted in light orange) and tidal breathing (depicted in green), which alternate under the guidance of physicians before breath-holding (depicted in yellow). The EIT electrodes are placed on the skin around the chest, resulting in the raw voltage signal containing changes from ventilation and perfusion inside the chest. To obtain accurate information about the ventilation activities, isolating the ventilation-related changes from the raw EIT voltage signal is necessary.

The heart beats at a significantly higher frequency of 60-100 times per minute (1-1.6 Hz) compared to the respiration rate of 10-20 times per minute (0.17-0.33 Hz). Furthermore, the magnitude of cardiac-related signals is much lower than that of ventilation-related signals. Therefore, digital filters are designed to effectively remove cardiac-related signals and noises, as illustrated in the lower row of Fig. \ref{raw}.

\subsection{EIT Image Reconstruction and Code Splicing}
The ventilation-related voltage signal is utilized as input to the image reconstruction algorithm (described in Section \ref{pre}), with the end of expiration selected as the reference point. Specifically, for each data record, a sequence of data frames at time $t_i$ (where $i=1,2,...,T$) denoted as $\bm d_i$ (where $i=1,2,...,T$) is reconstructed as a 3D image series $\bm P = {\bm p_i, (i=1,2,...,T)}$.

\begin{figure}[!h] 
	\centering
	\includegraphics[width=90 mm]{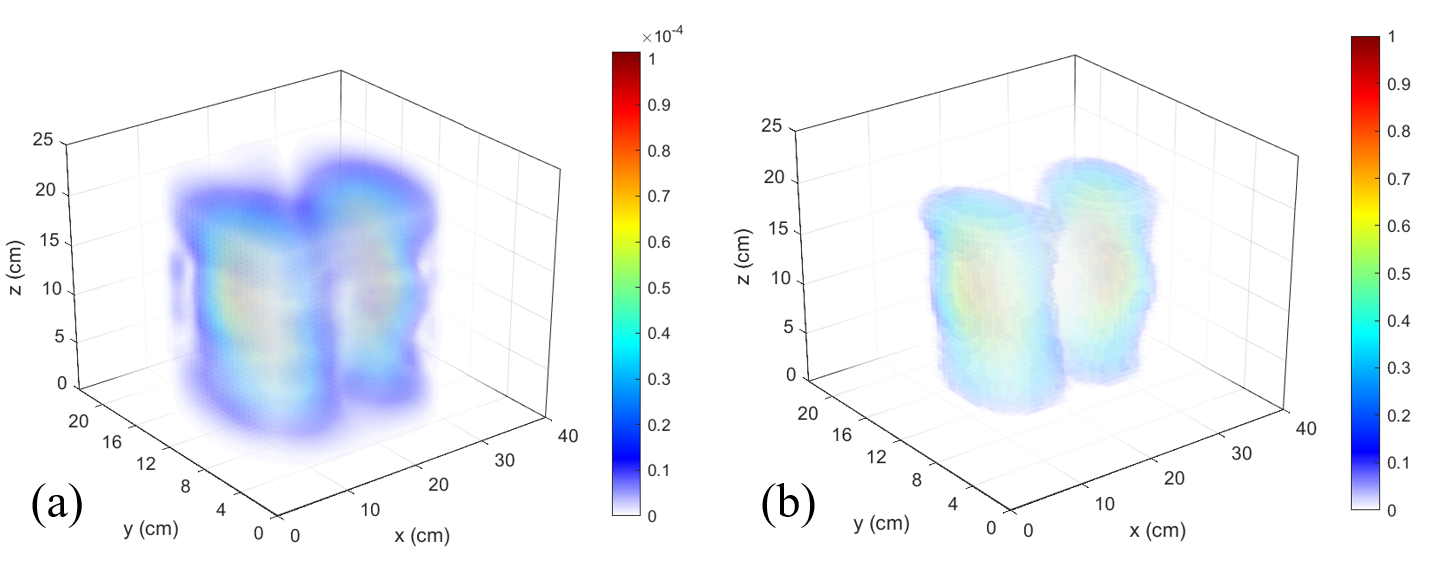}
	\caption{The EIT image (a) before and (b) after process. The unit in (a) is $mS/m$}
	\label{preprocess}
\end{figure}

To enhance the visualization of the lungs, the pixels that rank within the lowest 20\% are set to 0 for each image $\bm p_i$, which helps to make the lung outline more clearly visible. Subsequently, the amplitude of the entire image series $\bm P$ is normalized to fall within the $[0,1]$ range before being fed into the VAE. Let $P_{min}$ and $P_{max}$ denote the minimum and maximum pixel value of the image series $\bm P$, and the normalized image series $\hat {\bm P}$ can be expressed as shown in Eq.\ref{eq4}. The images before and after this process are presented in Fig.\ref{preprocess}.

\begin{equation}\label{eq4}
	\hat {\bm P} = \frac{\bm P - P_{min}}{P_{max} - P_{min}}
\end{equation}

\begin{figure*}[!h] 
	\centering
	\includegraphics[width = \linewidth]{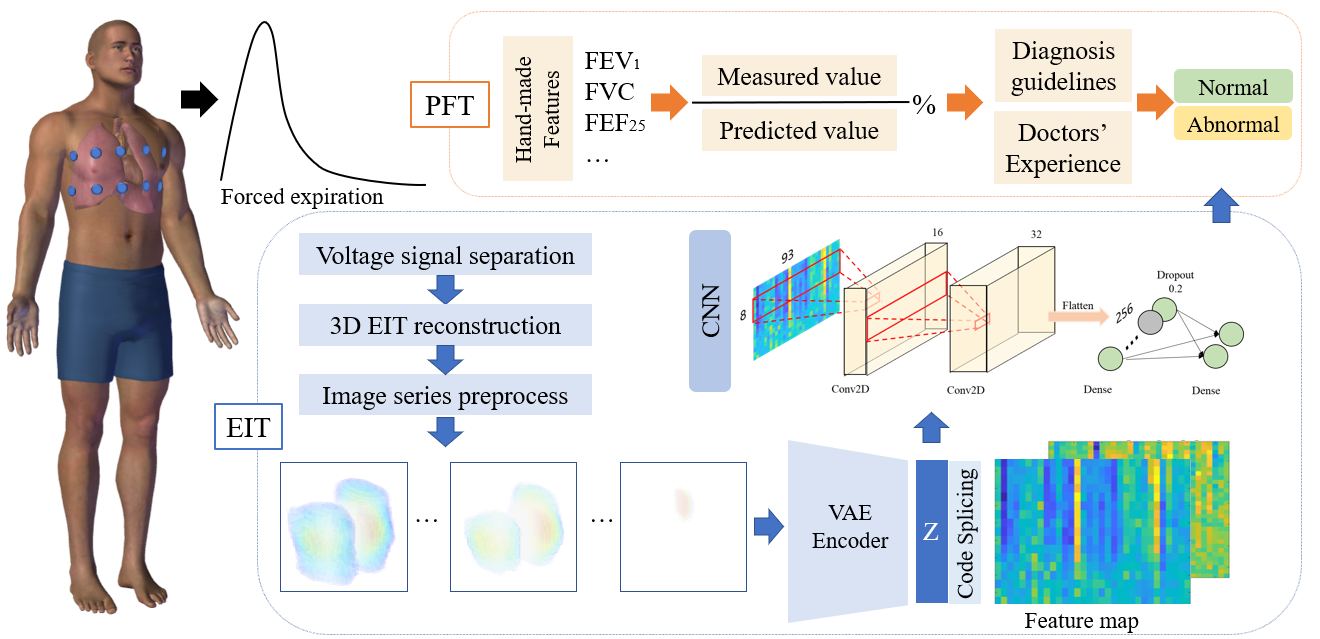}
	\caption{An overview of the proposed method. (The three-dimensional human body image is sourced from Zygote Media Group.)}
	\label{overview}
\end{figure*}

The trained VAE is utilized as a compressor for dimensional reduction. The resulting latent code of the reconstructed image series is represented as $\bm Z = {\bm z_{i}, (i=1,2,...,T)}$, where each $\bm z$ has a length of 32. Since the expiration duration varies from person to person, the $T$ dimension of the $\bm Z$ is zero-padded to a length of 93. The resulting zero-padded latent code is denoted as $\bm Z_{pad} = {\bm z_{i}, (i=1,2,...,93)}$. Subsequently, the $\bm Z_{pad}$ is input to a 2D CNN for classification. See Fig.\ref{overview} for a general overview of this work.

\section{EXPERIMENT DESIGN}
\label{exp}
\subsection{Subjects and Data Acquisition}
\subsubsection{Measurement system}
EIT signals were acquired using the Infivision 1900 (Beijing Huarui Boshi Medical Imaging Technology Co., Ltd., Beijing, China). Two electrode belts were placed around the chest to record the signals, with each belt containing 16 electrodes. The upper electrode belt was placed at the height of the armpit, while the lower electrode belt was placed at the fourth to sixth intercostal space (medioclavicular line).

The EIT measurement system utilized in this study has an input impedance of 40 k$\Omega$ at a phase angle of $-90^{\circ}$ and a frequency of 20 kHz. The instrumentation amplifier has a high standard mode rejection ratio (CMRR) of 120 dB. A 2-loop of electrodes injection-measurement pattern\cite{zhang2022deep} was used to record EIT signals, with an injected alternating current of 2 mA (root mean square) at a frequency of 20 kHz. EIT data were collected at a rate of 20 images per second and were reconstructed using a reconstruction matrix with Tikhonov regularization, as described in Section \ref{EIT}.

\subsubsection{Clinical Study Cohort}
From August 2021 to September 2022, a total of 186 subjects were recruited after obtaining written consent. Subjects who did not provide written informed consent (n=4), had contraindications to PFT or EIT (n=9), or had lung diseases such as pulmonary lesions, large bullae, and pleural effusion (n=11) were excluded prior to the test. During the test, 6 patients could not perform PFT adequately, and 19 had poor contact with EIT.

After the exclusions above, forced expiratory data from 137 subjects were included in the subsequent analysis. The participant's physical characteristics and PFT values are presented in Table \ref{infoSub}.
In general, 67 patients (age 62.36 ± 9.56 yr, body weight 63.15 ± 11.70 kg, body height 165.79 ± 8.04 cm) were classified as the abnormal ventilation group, while 70 patients (age 59.04 ± 10.37 yr, body weight 65.89 ± 11.53 kg, body height 166.60 ± 8.57 cm) were classified as the normal ventilation group. Demographically, no significant differences were observed between the two groups (P \textgreater  0.05).

EIT measurements were conducted with PFT with approval from the Ethics Committee of Zhongshan Hospital, Fudan University (2022-084R), and registration was in the Clinical Trials Register.

\begin{table}[!htbp]
	\caption{The physical characteristics of the participants}
	\includegraphics[width=90 mm]{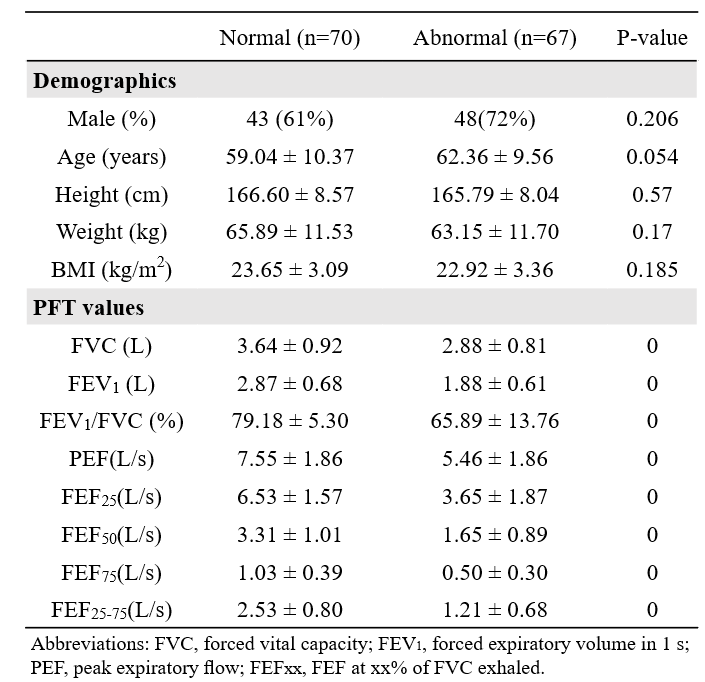}
	\label{infoSub}
\end{table}

\section{Results and Discussion}
\label{dis}
\subsection{Performance of VAE}
The EIT image series of the 137 subjects were processed and shuffled to create a training dataset for the VAE model, which was implemented and evaluated using Python with TensorFlow on an NVIDIA Tesla V100 GPU card. The image reconstruction was conducted using MATLAB R2021b. The dataset consisted of 2781 images with a size of $48 \times 32\times 48$, with 2508 images used for training and 279 for testing. The Adam optimizer was applied during training with a learning rate of $4 \times 10^{-4}$ over 50 epochs.

\begin{figure}[!htbp] 
	\centering
	\includegraphics[width = 90 mm]{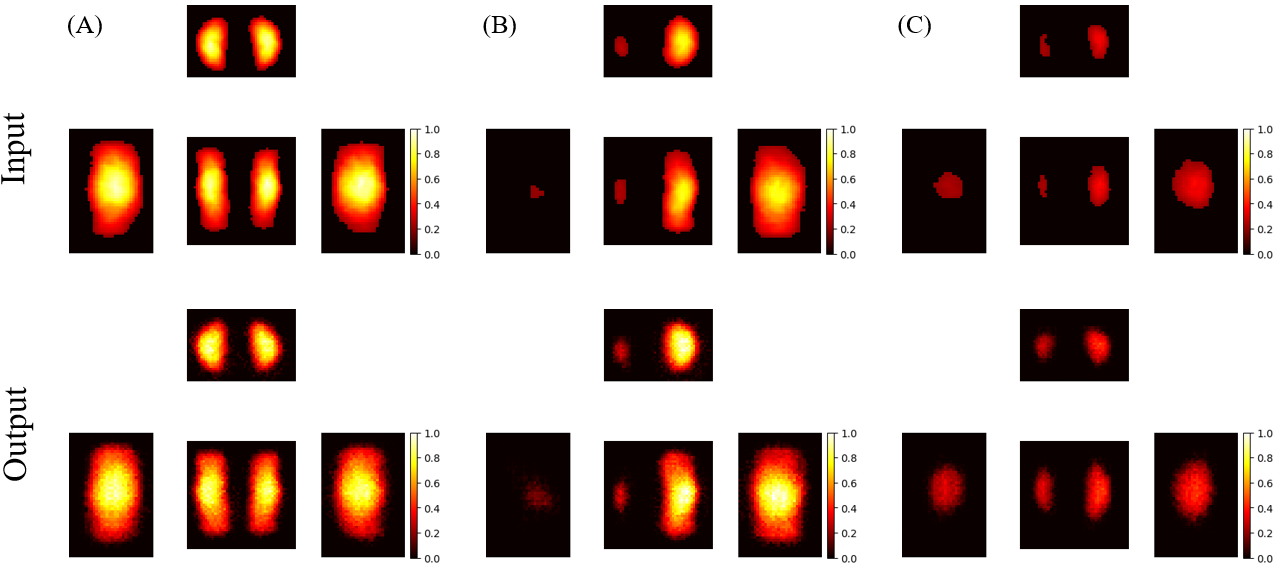}
	\caption{Three images reconstructed by the proposed VAE. The reconstructed images (bottom row) show good agreement with the original images (top row), demonstrating the high reconstruction accuracy of the VAE model.}
	\label{recon}
\end{figure}

The performance of the proposed method heavily relies on the effectiveness of VAE for dimensional reduction. To verify the reconstruction quality of VAE, several samples from the test dataset were randomly selected, and their reconstructed images were visually evaluated. As shown in Fig.\ref{recon}, each reconstructed 3D image is presented with four slices: the central coronal slice in the middle, the center section slice on top, and the left and right lung sections on both sides.

The upper row of Figure \ref{recon} shows the 3D EIT images of three subjects at different stages of the respiratory cycle. Figure A displays the image of a normal subject's lung at the apex of inhalation. The image shows both lungs as round and full, with the ventilation range of the left and right lungs being approximately the same. In contrast, Figure B shows the image of an abnormal subject at the start of forced expiration, with defects visible in the ventilation image of the right lung. Figure C shows the end of the expiration of another subject. The corresponding output images reconstructed by VAE are shown in the lower row of Figure \ref{recon}. The VAE effectively reproduces the input images in terms of both value and contour. 

\begin{figure}[!htbp] 
	\centering
	\includegraphics[width = 90mm]{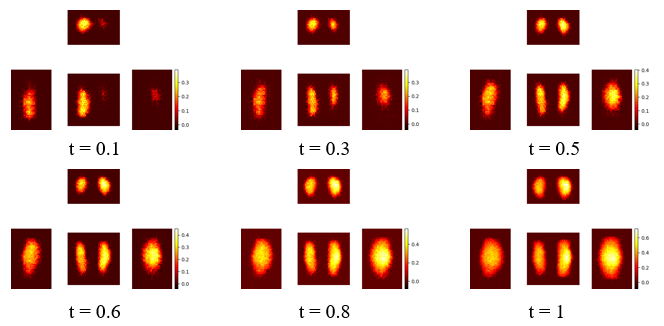}
	\caption{The intermediate images obtained by linearly interpolating between two latent vectors form a smooth transition, confirming the continuity of the latent space.}
	\label{连续性}
\end{figure}

Furthermore, to verify the compact and continuous nature of the latent space, we conducted code interpolation between two test samples $\bm x_1$ and $\bm x_2$. We first encode them to obtain their respective latent vectors $\bm z_1$ and $\bm z_2$. Then, we created a series of intermediate latent vectors by linearly interpolating between $\bm z_1$ and $\bm z_2$ through convex combinations of the two vectors, given by:

\begin{equation}
	\bm z_{i} = (1-t)\bm z_{1} + t\bm z_{2}
\end{equation}

where $t \in [0,1]$ is a parameter that controls the degree of interpolation. Subsequently, we decoded these intermediate vectors $\bm z_i$ to obtain the corresponding intermediate image $\bm x_i$. As depicted in Fig.\ref{连续性}, the intermediate images formed a smooth transition between the original images. Thus, we conclude that the latent code $\bm z$ provides a low-dimensional representation of the entire 3D image and that the latent space is compact and continuous. 

The latent code serves as a condensed representation of the spatial distribution of lung ventilation, enabling efficient storage and analysis of the pulmonary ventilation distribution's temporal changes. The sequence of latent space vectors, obtained from sequentially inputting 3D lung ventilation images, effectively captures the temporal changes in pulmonary ventilation distribution. Thus, this sequence of latent vectors can be utilized to efficiently store and analyze the temporal changes in the pulmonary ventilation distribution.

\subsection{Classification performance}
The input series were encoded and zero-padded to form a latent code series denoted as $\bm Z_{pad}$. As the training dataset consisted of 137 image sequences reconstructed from measured data, it was essential to validate the model to ensure that overfitting did not occur. To accomplish this, we employed ten-fold and leave-one-subject-out validation techniques during the convolutional neural network (CNN) training. Moreover, we recruited nine new subjects in October 2022, and their data were processed following the same pipeline as the training dataset. These data were used as blind data to test the model's generalization capability.

\subsubsection{Ten Fold and Leave-One-Out Cross Validation}
A ten-fold test is first conducted to ensure the model's fitness. The whole training data is split into 10 parts; each part is used as a test dataset in turn, while the remaining nine parts are used for training. The average accuracy and AUC are $0.956\pm 0.06$, and the f1-score is $0.956 \pm 0.06$. The coefficient of variation for the accuracy and AUC are $0.0637$ and $0.0639$, respectively, indicating a relatively low variance and good reproducibility of the results. These results suggest that the model has a good generalization ability and can accurately classify ventilation patterns in unseen data.

\begin{figure}[!h] 
	\centering
	\includegraphics[width = 90 mm]{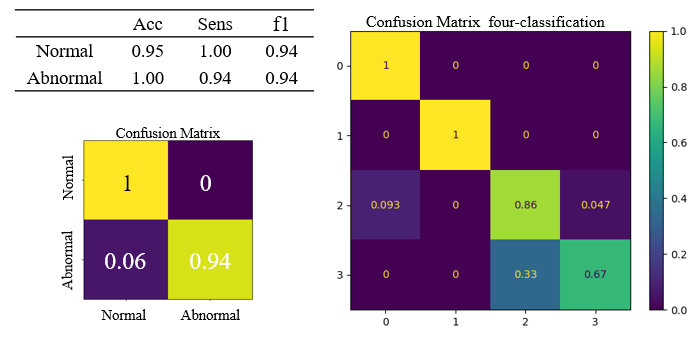}
	\caption{Results of Leave-one-out validation. (left: two-classification, right: four classifications, where 1,2,3,4 stands for normal, obstructed, restricted, and mixed, respectively.)}
	\label{LOO}
\end{figure}

Next, we performed Leave-One-Out Cross Validation (LOOCV) to evaluate the model's performance further. The accuracy, sensitivity, f1-score, and confusion matrix are shown in Fig.\ref{LOO}. The results indicate that the model achieves high accuracy, sensitivity, and f1-score on the test data. Specifically, the model achieves an overall accuracy of $0.953$, a sensitivity of $0.941$, and an f1-score of $0.945$. The confusion matrix shows that the model has a high true positive rate for all classes, indicating that the model can accurately classify the ventilation patterns for each subject. These results confirm the robustness and generalization ability of the proposed model for classifying the ventilation patterns in 3D lung ventilation images.

Furthermore, abnormal ventilation can be classified into obstructive, restrictive, and mixed patterns. In order to test the proposed pipeline, we modified the 2D CNN model from a two-class to a four-class classification. The LOOCV confusion matrix indicates good performance in identifying normal and obstructed patterns, where the obstructed pattern is characterized by slow and uneven expiration. However, the model's ability to distinguish between restrictive and mixed patterns could have been more satisfactory. Regarding respiratory mechanics, the restrictive pattern is characterized by a reduction in total capacity but a smooth expiration pattern similar to the normal pattern. The model's accuracy was lowest for the mixed pattern, mainly due to a lack of balanced training samples. Further investigation is required to verify and enhance the model's ability to differentiate among various lung ventilation abnormalities.

\subsubsection{Blind Data}
The proposed workflow was tested on blind data from nine patients with varying demographics and PFT results. Despite the unbalanced distribution in sex and lung ventilation mode, the results confirmed the reliability and validity of the proposed method. Among the nine subjects, only one was normal, and the remaining eight had obstructed or mixed diagnoses.

\begin{table}[!htbp]
	\caption{The physical characteristics of the $9$ participants}
	\includegraphics[width=90 mm]{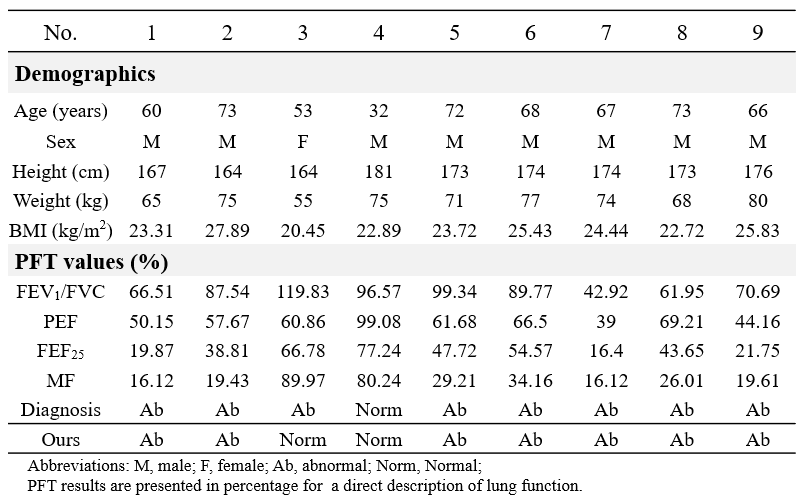} 
	\label{single9}
\end{table}

EIT records during forced expiration were processed using the proposed pipeline, and the PFT diagnosis was noted in Table \ref{single9}. The classification results of our method are shown in the last line of the table, with only one mistake in subject 3, a 53-year-old female with a PFT diagnosis of restriction. While the flow metrics in PFT, such as the percentage of $\text{FEV}_{1}/\text{FVC}$ and $\text{FEF}_{25-75}$ were close to normal, a decrease was observed in instantaneous flow rate, such as $\text{PEF}$ and $\text{FEF}_{25}$.

\section{Conclusion}
\label{conclu}
In summary, this work studies the general diagnosis of lung ventilation patterns using 3D EIT image series. Unlike previous studies focusing on specific diseases or operations, this work provides a more comprehensive diagnosis of normal or abnormal lung ventilation. Using a well-trained VAE network with MultiRes block, a single subject's spatial and temporal features are integrated into a two-dimensional feature map, which is then classified using a simple CNN network. The model exhibits satisfying accuracy and stability in cross-validation tests and is validated on new data from nine patients.

This study also addresses the need for more attention to individualized lung function assessment, which can provide valuable information for diagnosis and treatment. While PFT is commonly used for lung function diagnosis, the potential of EIT for individualized lung function evaluation is explored in this work. The results suggest that this approach may have promising applications in personalized diagnosis for lung function disorders.

While this work presents promising results for individualized lung function assessment using EIT, some limitations must be acknowledged. Firstly, the sample size of the training data set is relatively small, which may limit the generalizability of the proposed workflow. A larger sample size is needed to validate the results further and assess the model's performance among different populations.

Secondly, the study only focuses on forced expiration and does not consider the potential changes in lung function during normal breathing or other respiratory maneuvers. Incorporating more comprehensive respiratory measurements may provide a more complete assessment of lung function.

Finally, while the proposed workflow provides a two-dimensional feature map for classification, the interpretability of the features extracted by the VAE network and CNN still needs to be improved. Further research is needed to understand better the relationship between the extracted features and the underlying physiological mechanisms of lung function.

\bibliographystyle{unsrt}
\bibliography{references}

\end{document}